\documentclass{kapproc} 
\usepackage{procps}
\usepackage[dvips]{graphicx}

\upperandlowercase

\nochapequationnumber 

\let\footnote\savefootnote

\let\footnotetext\savefootnotetext

\let\footnoterule\savefootnoterule

\normallatexbib 

\begin{document}



\articletitle[Effective Lagrangians for QCD, Duality and Exact
Results]{Effective Lagrangians for QCD,\\~\\ Duality and Exact
Results}






 \author{Francesco Sannino}
 \affil{NORDITA, Blegdamsvej 17, Copenhagen, DK-2100, Denmark}
\email{francesco.sannino@nbi.dk}
















 \begin{abstract}
I briefly discuss effective Lagrangians for strong interactions
while concentrating on two specific lagrangians for QCD at large
matter density. I then introduce spectral duality in QCD a la
Montonen and Olive. The latter is already present in QCD in the
hadronic phase. However it becomes transparent at large chemical
potential. Finally I show the relevance of having exact non
perturbative constraints such as t'Hooft anomaly conditions at
zero and nonzero quark chemical potential on the possible phases
of strongly interacting matter. An important outcome is that for
three massless quarks at any chemical potential the only non
trivial solution of the constraints is chiral symmetry breaking.
This also shows that for three massless flavors at large quark
chemical potential CFL is the ground state.
\end{abstract}


\section{Effective Lagrangians for QCD}
In the non perturbative regime of strongly interacting theories
effective Lagrangians play a dominant role since they efficiently
describe the non perturbative dynamics in terms of the relevant
degrees of freedom. Symmetries, anomalous and exact, are used to
constrain the effective Lagrangians. An important point is that
the effective Lagrangian approach is applicable to any region of
the QCD or QCD-like phase diagram whenever the relevant degrees of
freedom and the associated symmetries are defined.

\subsection{Zero temperature and quark chemical potential}
 At zero temperature and quark chemical potential the simplest
effective Lagrangian describing a relevant part of the
nonperturbative physics of the Yang-Mills (YM) theory is the
glueball Lagrangian whose potential is
\cite{schechter,{joe},{MS},{SST}}:
\begin{eqnarray} V=\frac{H}{2}\ln \left[\frac{H}{\Lambda^4}\right] \ . \end{eqnarray}
The latter is constrained using trace anomaly and $H\sim {\rm
Tr}\left[G^{\mu\nu}G_{\mu\nu}\right]$ with $G^{\mu\nu}$ the gluon
field stress tensor. It describes the vacuum of a generic
Yang-Mills theory. A similar effective Lagrangian (using
superconformal anomalies) can be written for the non perturbative
super Yang-Mills (SYM) theory. This is the celebrated
Veneziano-Yankielowicz \cite{Veneziano:1982ah} lagrangian.  In
Yang Mills and super Yang-Mills theories no exact continuous
global symmetries are present which can break spontaneously and
hence no goldstones are present. The situation is different when
flavors are included in the theory. Here the spontaneous breaking
of chiral symmetry leads to a large number of goldstone's
excitations. We note that in
\cite{Marotta:2002gc,{Marotta:2002ns}} we were able, using string
techniques, to derive a number of fundamental perturbative and non
perturbative properties for supersymmetric QCD such as the beta
function, fermion condensate as well as chiral anomalies.

Recently in \cite{Sannino:2003xe} we constructed effective
Lagrangians of the Veneziano-Yankielowicz (VY) type for two
non-supersymmetric but strongly interacting theories with a dirac
fermion either in the two index symmetric or two index
antisymmetrix representation of the gauge group. These theories
are planar equivalent, at $N\to\infty$ to SYM
\cite{Armoni:2003gp}. In this limit the non-supersymmetric
effective Lagrangians coincide with the bosonic part of the VY
Lagrangian.

We departed from the supersymmetric limit in two ways. First, we
considered finite values of $N$. Then $1/N$ effects break
supersymmetry. We suggested the simplest modification of the VY
Lagrangian which incorporates these $1/N$ effects, leading to a
non-vanishing vacuum energy density. We analyzed the spectrum of
the finite-$N$ non-supersymmetric daughters. For $N=3$ the
two-index antisymmetric representation (one flavor) {\it is
one-flavor QCD}. We showed that in this case the scalar
quark-antiquark state is heavier than the corresponding
pseudoscalar state, the $\eta^{\prime}$. Second, we added a small
fermion mass term which breaks supersymmetry explicitly. The
vacuum degeneracy is lifted, the parity doublets split and we
evaluated this splitting. The $\theta$-angle dependence and its
implications were also investigated. This new effective Lagrangian
provides a number of fundamental results about QCD which can be
already tested either experimentally or via lattice simulations.

This new type of expansion in the inverse of number colors in
which the quark representation is the two index antisymmetric
representation of the gauge group at any given $N$ may very well
be more convergent then the ordinary $1/N$ expansion. In the
ordinary case one keeps the fermion in the fundamental
representation of the gauge group while increasing the number of
colors. Indeed recently in \cite{Harada:2003em} we have studied
the dependence on the number of colors (while keeping the fermions
in the fundamental representation of the gauge group) of the
leading pi pi scattering amplitude in chiral dynamics. We have
demonstrated the existence of a critical number of colors for and
above which the low energy pi pi scattering amplitude computed
from the simple sum of the current algebra and vector meson terms
is crossing symmetric and unitary at leading order in a $1/N$
expansion. The critical number of colors turns out to be $N=6$ and
is insensitive to the explicit breaking of chiral symmetry. This
means that the ordinary $1/N$ corrections for the real world are
large.

\subsection{Nonzero temperature and quark chemical potential}

At nonzero temperature the center of the $SU(N)$ gauge group
becomes a relevant symmetry \cite{Svetitsky:1982gs}. However
except for mathematically defined objects such as Polyakov loops
the physical states of the theory are neutral under the center
group symmetry.

A new class of effective Lagrangians have been constructed to show
how the information about the center group symmetry is efficiently
transferred to the actual physical states of the theory
\cite{Sannino:2002wb,{Mocsy:2003tr},{Mocsy:2003un},{Mocsy:2003qw}}
and will be reviewed in detail elsewhere. Via these Lagrangians we
were also able to have a deeper understanding of the relation
between chiral restoration and deconfinement \cite{Mocsy:2003qw}
for quarks in the fundamental and in the adjoint representation of
the gauge group.

I will focus here on the two basic effective Lagrangians developed
for color superconductivity. More specifically the lagrangian for
the color flavor locked phase (CFL) of QCD at high chemical
potential and the 2 flavor color superconductive effective
Lagrangian.

\label{uno} A color superconducting phase is a reasonable
candidate for the state of strongly interacting matter for very
large quark chemical potential
\cite{{Barrois:1977xd},{Bar_79},{Bailin:1984bm},{Alford:1998zt},{Rapp:1998zu}}.
Many properties of such a state have been investigated for two and
three flavor QCD. In some cases these results rely heavily on
perturbation theory, which is applicable for very large chemical
potentials. Some initial applications to supernovae explosions and
gamma ray bursts can be found in \cite{HHS} and
\cite{Ouyed:2001cg} respectively, see also \cite{Blaschke:2003yn}.
{}The interested reader can find a discussion of the effects of
color superconductivity on the mass-radius relationship of compact
stars in \cite{Alford:2003eg}

\section{Color Flavor Locked Phase} \label{3f}

{}For $N_{f}=3$ light flavors at very high chemical potential
dynamical computations suggest that the preferred phase is a
superconductive one and the following ansatz for a quark-quark
type of condensate is energetically favored:
\begin{equation}
\epsilon ^{\alpha \beta }<q_{L\alpha ;a,i}q_{L\beta ;b,j}>\sim
k_{1}\delta _{ai}\delta _{bj}+k_{2}\delta _{aj}\delta _{bi}\ .
\label{condensate}
\end{equation}
\noindent A similar expression holds for the right transforming
fields. The Greek indices represent spin, $a$ and $b$ denote color
while $i$ and $j$ indicate flavor. The condensate breaks the gauge
group completely while locking the left/right transformations with
color. The final global symmetry group is $SU_{c+L+R}(3)$, and the
low energy spectrum consists of $9$ Goldstone bosons.

\section{Duality made transparent in QCD}

Here we seek insight regarding the relevant energy scales of
various physical states of the color flavor locked phase (CFL),
such as the vector mesons and the solitons \cite{Jackson:2003dk}.
Our results do not support the naive expectation that all massive
states are of the order of the color superconductive gap,
$\Delta$. Our strategy is based on exploiting the significant
information already contained in the low--energy effective theory
for the massless states. We transfer this information to the
massive states of the theory by making use of the fact that higher
derivative operators in the low--energy effective theory for the
lightest state can also be induced when integrating out heavy
fields. {}For the vector mesons, this can be seen by considering a
generic theory containing vector mesons and Goldstone bosons.
After integrating out the vector mesons, the induced local
effective Lagrangian terms for the Goldstone bosons must match the
local contact terms from operator counting. We find that each
derivative in the (CFL) chiral expansion is replaced by a vector
field $\rho_{\mu}$ as follows
\begin{eqnarray}
\partial \rightarrow \frac{\Delta}{F_{\pi}}\rho\ . \end{eqnarray}
This relation allows us to deduce, among other things, that the
energy scale for the vector mesons is
\begin{eqnarray}
m_v \sim \Delta \ ,
\end{eqnarray}
where $m_v$ is the vector meson mass. Our result is in agreement
with the findings in \cite{{Casalbuoni:2000na},{Rho:2000ww}}. We
shall see that this also suggests that the KSRF relation holds in
the CFL phase.

{}In the solitonic sector, the CFL chiral Lagrangian
\cite{Hong:1999dk,{CG}} gives us the scaling behavior of the
coefficient of the Skyrme term and thus shows that the mass of the
soliton is of the order of
\begin{eqnarray}
M_{\rm soliton} \sim \frac{F^2_{\pi}}{\Delta} \ ,
\end{eqnarray}
which is contrary to naive expectations. This is suggestive of a
kind of duality between vector mesons and solitons in the same
spirit as the duality advocated some years ago by Montonen and
Olive for the $SU(2)$ Georgi-Glashow theory
\cite{Montonen:1977sn}. This duality becomes more apparent when
considering the product
\begin{eqnarray}
M_{\rm soliton} m_v \sim F^2_{\pi} \ ,
\end{eqnarray}
which is independent of the scale, $\Delta$. In the present case,
if the vector meson self-coupling is $\widetilde{g}$, we find that
the Skyrme coefficient, $e\sim \Delta/F_{\pi}$, can be identified
with $\widetilde{g}$.  Thus, the following relations hold:
\begin{eqnarray}
M_{\rm soliton} \propto \,\frac{F_{\pi}}{\widetilde{g}}  \quad
{\rm and}\quad m_{v}\propto \,\widetilde{g}\, F_{\pi} \ .
\end{eqnarray}
In this notation the electric-magnetic (i.e. vector meson-soliton)
duality is transparent. Since the topological Wess-Zumino term in
the CFL phase is identical to that in vacuum, we identify the
soliton with a physical state having the quantum numbers of the
nucleon. We expect that the product of the nucleon and vector
meson masses will scale like $F_{\pi}^2$ for any non-zero chemical
potential for three flavors. Interestingly, quark-hadron
continuity can be related to duality \cite{Schafer:1998ef}.
Testing this relation can also be understood as a quantitative
check of quark-hadron continuity. It is important to note that our
results are tree level results and that the resulting duality
relation can be affected by quantum corrections. Our results have
direct phenomenological consequences for the physics of compact
stars with a CFL phase.  While vector mesons are expected to play
a relevant role, solitons can safely be neglected for large values
of the quark chemical potential.

\subsection{The Lagrangian for CFL Goldstones} When diquarks
condense for the three flavor case, we have the following symmetry
breaking:
\begin{eqnarray}
\left[SU_c(3)\right] \times SU_L(3) \times SU_R(3) \times U_B(1)
\rightarrow SU_{c+L+R}(3) \ . \nonumber
\end{eqnarray}
The gauge group undergoes a dynamical Higgs mechanism, and nine
Goldstone bosons emerge. Neglecting the Goldstone mode associated
with the baryon number and quark masses (which will not be
important for our discussion at lowest order), the derivative
expansion of the effective Lagrangian describing the octect of
Goldstone bosons is \cite{Hong:1999dk,{CG}}:
\begin{eqnarray}
{\cal L}=\frac{F^2_{\pi}}{8}{\rm Tr}\left[\partial_{\mu}U
\partial^{\mu}U^{\dagger}\right]\equiv \frac{F^2_{\pi}}{2}
{\rm Tr}\left[p_{\mu}p^{\mu}\right] \ ,
\end{eqnarray}
with $p_{\mu}=\frac{i}{2} \left(\xi
\partial_{\mu}\xi^{\dagger} -
\xi^{\dagger}\partial_{\mu}\xi\right)$,  $U=\xi ^2$,
 $\xi=e^{i\frac{\phi}{F_{\pi}}}$ and $\phi$ is the octet of
Goldstone bosons. $U$ transforms linearly according to $g_L U
g_R^{\dagger}$  and $g_{L/R}\in SU_{L/R}(3)$ while $\xi$
transforms non-linearly:
\begin{eqnarray}
\xi \rightarrow g_L\,\xi \,K^{\dagger}\left(\phi,
g_L,g_R\right)\equiv K\left(\phi, g_L,g_R\right)\,\xi\,
g_R^{\dagger} \ .
\end{eqnarray}
This constraint implicitly defines the matrix, $K\left(\phi,
g_L,g_R\right)$.   Here, we wish to examine the CFL spectrum of
massive states using the technique of integrating in/out at the
level of the effective Lagrangian. $F_{\pi}$ is the Goldstone
boson decay constant.  It is a non-perturbative quantity whose
value is determined experimentally or by non-perturbative
techniques (e.g.\ lattice computation). For very large quark
chemical potential, $F_{\pi}$ can be estimated perturbatively. It
is found to be proportional to the Fermi momentum, $p_F\sim \mu$,
with $\mu$ the quark chemical potential \cite{Schafer:2003vz}.
Since a frame must be fixed in order to introduce a chemical
potential, spatial and temporal components of the effective
Lagrangians split. This point, however, is not relevant for the
validity of our results.

When going beyond the lowest-order term in derivatives, we need a
counting scheme.  For theories with only one relevant scale (such
as QCD at zero chemical potential), each derivative is suppressed
by a factor of $F_{\pi}$.  This is not the case for theories with
multiple scales. In the CFL phase, we have both $F_{\pi}$ and the
gap, $\Delta$, and the general form of the chiral expansion is
\cite{Schafer:2003vz}:
\begin{eqnarray}
L\sim F^2_{\pi}\Delta^2
\left(\frac{\vec{\partial}}{\Delta}\right)^{k}
\left(\frac{\partial_0}{\Delta}\right)^{l} U^{m} {U^{\dagger}}^{n}
\ .
\end{eqnarray}
Following \cite{Schafer:2003vz}, we distinguish between temporal
and spatial derivatives.  Chiral loops are suppressed by powers of
$p/4\pi F_{\pi}$, and higher-order contact terms are suppressed by
$p/\Delta$ where $p$ is the momentum. Thus, chiral loops are
parametrically small compared to contact terms when the chemical
potential is large.

There is also a topological term which is essential in order to
satisfy the t'Hooft anomaly conditions
\cite{S,{HSaS},{Sannino:2003pq}} at the effective Lagrangian
level.  It is important to note that respecting the t'Hooft
anomaly conditions is more than an academic exercise. In fact, it
requires that the form of the Wess-Zumino term is the same in
vacuum and at non-zero chemical potential. Its real importance
lies in the fact that it forbids a number of otherwise allowed
phases which cannot be ruled out given our rudimentary treatment
of the non-perturbative physics.  As an example, consider a phase
with massless protons and neutrons in three-color QCD with three
flavors.  In this case chiral symmetry does not break. This is a
reasonable realization of QCD for any chemical potential. However,
it does not satisfy the t'Hooft anomaly conditions and hence
cannot be considered.  Were it not for the t'Hooft anomaly
conditions, such a phase could compete with the CFL phase.

Gauging the Wess-Zumino term with to respect the electromagnetic
interactions yields the familiar $\pi^0\rightarrow 2\gamma$
anomalous decay. This term \cite{WZ} can be written compactly
using the language of differential forms. It is useful to
introduce the algebra-valued Maurer-Cartan one form $\alpha
=\alpha_{\mu}dx^{\mu}=\left( \partial _{\mu }U\right)
U^{-1}\,dx^{\mu }\equiv \left( dU\right) U^{-1}$
which transforms only under the left $SU_{L}(3)$ flavor group. The
Wess-Zumino effective action is
\begin{eqnarray}
\Gamma _{WZ}\left[ U\right] =C\,\int_{M^{5}}{\rm Tr}\left[ \alpha
^{5}\right] \ .  \label{WZ}
\end{eqnarray}
The price which must be paid in order to make the action local is
that the spatial dimension must be augmented by one.  Hence, the
integral must be performed over a five-dimensional manifold whose
boundary ($M^{4}$) is ordinary Minkowski space. In
\cite{{Hong:1999dk},{S},Casalbuoni:2000jn} the constant $C$ has
been shown to be the same as that at zero density, i.e.
\begin{equation}
C=-i\frac{N_{c}}{240\pi ^{2}}\ , \label{coef}
\end{equation}
where $N_{c}$ is the number of colors (three in this case).  Due
to the topological nature of the Wess-Zumino term its coefficient
is a pure number.

\subsection{The vector mesons} {}It is well known that massive
states are relevant for low energy dynamics.  Consider, for
example, the role played by vector mesons in pion-pion scattering
\cite{{Harada:2003em},Sannino:1995ik} in saturating the unitarity
bounds. More specifically, vector mesons play a relevant role when
describing the low energy phenomenology of QCD and may also play a
role also in the dynamics of compact stars with a CFL core
\cite{Vogt:2003ph}. In order to investigate the effects of such
states, we need to know their in--medium properties including
their gaps and the strength of their couplings to the CFL
Goldstone bosons. Except for the extra spontaneously broken
$U(1)_B$ symmetry, the symmetry properties of the CFL phase have
much in common with those of zero density phase of QCD.  This fact
allows us to make some non perturbative but reasonable estimates
of vector mesons properties in medium.  We have already presented
the general form of the chiral expansion in the CFL phase.  As
will soon become clear, we are now interested in the
four--derivative (non--topological) terms whose coefficients are
proportional to
\begin{eqnarray}
\frac{F^2_{\pi}}{\Delta^2} \ .
\end{eqnarray}
This must be contrasted with the situation at zero chemical
potential, where the coefficient of the four--derivative term is
always a pure number before quantum corrections are taken into
account. In vacuum, the tree-level Lagrangian which simultaneously
describes vector mesons, Goldstone bosons, and their interactions
is:
\begin{eqnarray}
L&=&\frac{F^2_{\pi}}{2}{\rm
Tr}\left[p_{\mu}p^{\mu}\right]+\frac{m^2_{v}}{2} {\rm Tr}
\left[\left(\rho_{\mu} +
\frac{v_{\mu}}{\widetilde{g}}\right)^2\right] \nonumber \\
&-&\frac{1}{4}{\rm Tr}
\left[F_{\mu\nu}(\rho)F^{\mu\nu}(\rho)\right] \ ,
\end{eqnarray}
where $F_{\pi}\simeq 132$ Mev and $v_{\mu}$ is the one form
$v_{\mu}=\frac{i}{2} \left(\xi \partial_{\mu}\xi^{\dagger} +
\xi^{\dagger}\partial_{\mu}\xi\right)$
with $U=\xi^2$ and $F_{\mu\nu}(\rho)=\partial_{\mu}\rho_{\nu} -
\partial_{\nu}\rho_{\mu} + i \widetilde{g}\,
[\rho_{\mu},\rho_{\nu}]$. At tree level this Lagrangian agrees
with the hidden local symmetry results \cite{Bando:1987br}.

When the vector mesons are very heavy with respect to relevant
momenta, they can be integrated out. This results in the field
constraint:
\begin{eqnarray}
\rho_{\mu} = -\frac{v_{\mu}}{\widetilde{g}} \ .
\end{eqnarray}
Substitution of this relation in the vector meson kinetic term
(i.e., the replacement of $F_{\mu\nu}(\rho)$ by $F_{\mu\nu}(v)$)
gives the following four derivative operator with two time
derivatives and two space derivatives \cite{Schechter:1999hg}:
\begin{eqnarray}  \frac{1}{64\,\widetilde{g}^2}{\rm Tr}\left[[\alpha_{\mu}
,\alpha_{\nu}]^2\right] \ .\label{4d}\end{eqnarray} The
coefficient is proportional to $1/\widetilde{g}^2$. It is also
relevant to note that since we are describing physical fields we
have considered canonically normalized fields and kinetic terms.
This Lagrangian can also be applied to the CFL case. In the
vacuum, $\widetilde{g}$ is a number of order one independent of
the scale at tree level. This is no longer the case in the CFL
phase. Here, by comparing the coefficient of the four--derivative
operator in eq.~(\ref{4d}) obtained after having integrated out
the vector meson with the coefficient of the same operator in the
CFL chiral perturbation theory we determine the following scaling
behavior of $\widetilde{g}$:
\begin{eqnarray}
\widetilde{g} \propto \frac{\Delta}{F_{\pi}} \ . \label{first}
\end{eqnarray}
By expanding the effective Lagrangian with the respect to the
Goldstone boson fields, one sees that $\widetilde{g}$ is also
connected to the vector meson coupling to two pions, $g_{\rho \pi
\pi}$, through the relation
\begin{eqnarray}
g_{\rho\pi \pi} = \frac{m^2_{v}}{\widetilde{g}F^2_{\pi}} \ .
\end{eqnarray}
In vacuum $g_{\rho\pi \pi } \simeq 8.56$ and $\widetilde{g} \simeq
3.96$ are quantities of order one. Since $v_{\mu}$ is essentially
a single derivative, the scaling behavior of $\widetilde{g}$
allows us to conclude that each derivative term is equivalent to
$\widetilde{g}\,\rho_{\mu}$ with respect to the chiral expansion.
For example, dropping the dimensionless field $U$, the operator
with two derivatives becomes a mass operator for the vector meson
\begin{eqnarray}
F^2_{\pi} \partial^2_{\mu} \rightarrow F_{\pi}^2 \,
\widetilde{g}^2\, \rho^2_{\mu} \sim \Delta^2 \rho^2_{\mu}\ .
\end{eqnarray}
This demonstrates that the vector meson mass gap is proportional
to the color superconducting gap. This non-perturbative result is
relevant for phenomenological applications. It is interesting to
note that our simple counting argument agrees with the underlying
QCD perturbative computations of Ref.~\cite{Casalbuoni:2000na} and
also with recent results of Ref.~\cite{Rho:2000ww}. In
\cite{Manuel:2000xt}, at high chemical potential, vector meson
dominance is discussed. However, our approach is more general
since it does not rely on any underlying perturbation theory. It
can be applied to theories with multiple scales for which the
counting of the Goldstone modes is known. Since $m^2_v \sim
\Delta^2$, we find that $g_{\rho\pi\pi}$ scales with
$\widetilde{g}$ suggesting that the KSRF relation is a good
approximation also in the CFL phase of QCD.

\subsection{CFL-Solitons} The low energy effective theory supports
solitonic excitations which can be identified with the baryonic
sector of the theory at non-zero chemical potential. In order to
obtain classically stable configurations, it is necessary to
include at least a four--derivative term (containing two temporal
derivatives) in addition to the usual two--derivative term. Such a
term is the Skyrme term:
\begin{eqnarray}
L^{\rm skyrme}=\frac{1}{32\,e^2}{\rm Tr}\left[[\alpha_{\mu}
,\alpha_{\nu}]^2\right] \ .
\end{eqnarray}
Since this is a fourth--order term in derivatives not associated
with the topological term we have:
\begin{eqnarray}
e \sim \frac{\Delta}{ F _{\pi}} \ .
\end{eqnarray}
This term is the same as that which emerges after integrating out
the vector mesons (see eq.~(\ref{4d})), and one concludes that $e
=\sqrt{2}\,\widetilde{g}$ \cite{Schechter:1999hg}. The simplest
complete action supporting solitonic excitations is:
\begin{eqnarray}
\int\,d^4x\left[\frac{F^2_{\pi}}{2}{\rm
Tr}\left[p_{\mu}p^{\mu}\right] + L^{\rm skyrme}\right] +
\Gamma_{WZ} \ .
\end{eqnarray}
The Wess-Zumino term in eq.~(\ref{WZ}) guarantees the correct
quantization of the soliton as a spin $1/2$ object. Here we
neglect the breaking of Lorentz symmetries, irrelevant to our
discussion. The Euler-Lagrangian equations of motion for the
classical, time independent, chiral field $U_0(\bf{r})$ are highly
non-linear partial differential equations. To simplify these
equations Skyrme adopted the hedgehog {\it ansatz} which, suitably
generalized for the three flavor case, reads
\cite{Schechter:1999hg}: \begin{eqnarray} U_0(\bf{r}) =\left(%
\begin{array}{cc}
   e^{i {\vec{\tau}}\cdot \hat{r} F(r)}& 0 \\
  0& 1 \\
\end{array}%
\right)\ ,
\end{eqnarray}
where $\vec{\tau}$ represents the Pauli matrices and the radial
function $F(r)$ is called the chiral angle. The {\it ansatz} is
supplemented with the boundary conditions $F(\infty)=0$ and
$F(0)=0$ which guarantee that the configuration posseses unit
baryon number. After substituting the {\it ansatz} in the action
one finds that the classical solitonic mass is, up to a numerical
factor:
\begin{eqnarray}
M_{\rm soliton} \propto \frac{F_{\pi}}{e} \sim
\frac{F^2_{\pi}}{\Delta}\ ,
\end{eqnarray}
and the isoscalar radius, $\langle r^2\rangle_{I=0}\sim
1/({F^2_{\pi}\,e^2})\sim 1/\Delta^2$. Interestingly, due to the
non perturbative nature of the soliton, its mass turns to be dual
to the vector meson mass. It is also clear that although the
vector mesons and the solitons have dual masses, they describe two
very distinct types of states. The present duality is very similar
to the one argued in \cite{Montonen:1977sn}. Indeed, after
introducing the collective coordinate quantization, the soliton
(due to the Wess-Zumino term) describes baryonic states of
half-integer spin while the vectors are spin one mesons. Here, the
dual nature of the soliton with respect to the vector meson is
enhanced by the fact that, in the CFL state, $\widetilde{g}\sim
\Delta/F_{\pi}$ is expected to be substantially reduced with
respect to its value in vacuum. Once the soliton is identified
with the nucleon (whose density--dependent mass is denoted with
$M_N(\mu)$) we predict the following relation to be independent of
the matter density:
\begin{eqnarray}
\frac{M_{N}(\mu)\,m_{v}(\mu)}{(2\pi F_{\pi}(\mu))^2} =
\frac{M_{N}(0)\,m_{v}(0)}{(2\pi F_{\pi}(0))^2}\sim 1.05 \ .
\end{eqnarray}
In this way, we can relate duality to quark-hadron continuity. We
considered duality, which is already present at zero chemical
potential, between the soliton and the vector mesons a fundamental
property of the spectrum of QCD which should persists as we
increase the quark chemical potential. Should be noted that
differently than in \cite{Hong:2000ff} we have not subtracted the
energy cost to excite a soliton from the fermi sea. Since we are
already considering the Lagrangian written for the excitations
near the fermi surface we would expect not to consider such a
corrections. In any event this is of the order $\mu$
\cite{Hong:2000ff} and hence negligible with respect to $M_{\rm
soliton}$.

We have shown that the vector mesons in the CFL phase have masses
of the order of the color superconductive gap, $\Delta$. On the
other hand the solitons have masses proportional to
$F^2_{\pi}/\Delta$ and hence should play no role for the physics
of the CFL phase at large chemical potential. We have noted that
the product of the soliton mass and the vector meson mass is
independent of the gap. This behavior reflects a form of
electromagnetic duality in the sense of Montonen and Olive
\cite{Montonen:1977sn}. We have predicted that the nucleon mass
times the vector meson mass scales as the square of the pion decay
constant at any nonzero chemical potential. In the presence of two
or more scales provided by the underlying theory the spectrum of
massive states shows very different behaviors which cannot be
obtained by assuming a naive dimensional analysis.

\section{2 SC General Features and Effective Lagrangian} \label{tre}

QCD with 2 massless flavors has gauge symmetry $SU_{c}(3)$ and
global symmetry
\begin{equation}
SU_{L}(2)\times SU_{R}(2)\times U_{V}(1)\ .
\end{equation}
At very high quark density the ordinary Goldstone phase is no
longer favored compared with a superconductive one associated to
the following type of diquark condensates:
\begin{equation}
\langle L{^{\dagger }}^{a}\rangle \sim \langle \epsilon
^{abc}\epsilon ^{ij}q_{Lb,i}^{\alpha }q_{Lc,j;\alpha }\rangle \
,\qquad \langle R{^{\dagger }}^{a}\rangle \sim -\langle \epsilon
^{abc}\epsilon ^{ij}q_{Rb,i;\dot{\alpha}
}q_{Rc,j}^{\dot{\alpha}}\rangle \ ,
\end{equation}
If parity is not broken spontaneously, we have $\left\langle
L_{a}\right\rangle =\left\langle R_{a}\right\rangle =f\delta
_{a}^{3}$, where we choose the condensate to be in the 3rd
direction of color. The order parameters are singlets under the
$SU_{L}(2)\times SU_{R}(2)$ flavor transformations while
possessing baryon charge $\frac{2}{3}$. The vev leaves invariant
the following symmetry group:
\begin{equation}
\left[ SU_{c}(2)\right] \times SU_{L}(2)\times SU_{R}(2)\times
\widetilde{U}_{V}(1)\ ,
\end{equation}
where $\left[ SU_{c}(2)\right] $ is the unbroken part of the gauge
group. The $\widetilde{U}_{V}(1)$ generator $\widetilde{B}$ is the
following linear combination of the previous $U_{V}(1)$ generator
$B$ and the broken diagonal generator of the $SU_{c}(3)$ gauge
group $T^{8}$:
$\displaystyle{\widetilde{B}=B-\frac{2\sqrt{3}}{3}T^{8}={\rm
diag}(0,0,1)}$. The quarks with color $1$ and $2$ are neutral
under $\widetilde{B}$ and consequently so is the condensate.


The spectrum in the 2SC state is made of 5 massive Gluons with a
mass of the order of the gap, 3 massless Gluons confined (at zero
temperature) into light glueballs and gapless up and down quarks
in the direction (say) 3 of color. \subsection{The 5 massive
Gluons} The relevant coset space $G/H$
\cite{CDS,{Casalbuoni:2000jn}} with
\begin{eqnarray}
G=SU_{c}(3)\times U_{V}(1)\ , \quad   {\rm and} \quad
H=SU_{c}(2)\times \widetilde{U}_{V}(1) \end{eqnarray} is
parameterized by:
\begin{eqnarray}
{V}=\exp (i\xi ^{i}X^{i})\ ,
\end{eqnarray}
where $\{X^{i}\}$ $i=1,\cdots ,5$ belong to the coset space $G/H$
and are taken to be $X^{i}=T^{i+3}$ for $i=1,\cdots ,4$ while
\begin{eqnarray}
X^{5}=B+\frac{\sqrt{3}}{3}T^{8}={\rm
diag}(\frac{1}{2},\frac{1}{2},0)\ . \label{broken}
\end{eqnarray}
$T^{a}$ are the standard generators of $SU(3)$. The coordinates
\begin{eqnarray} \nonumber
\xi ^{i}=\frac{\Pi ^{i}}{f}\quad i=1,2,3,4\ ,\qquad \xi
^{5}=\frac{\Pi ^{5}}{\widetilde{f}}\ ,
\end{eqnarray}
via $\Pi $ describe the Goldstone bosons which will be absorbed in
the longitudinal components of the gluons. The vevs $f$ and
$\widetilde{f}$ are, at asymptotically high densities,
proportional to $\mu $. ${V}$ transforms non linearly:
\begin{eqnarray}
{V}(\xi )\rightarrow u_{V}\,g \,{V}(\xi )\,h^{\dagger }(\xi
,g,u)\,h_{\widetilde{V}}^{\dagger }(\xi ,g,u)\ , \label{nl2}
\end{eqnarray}
with
\begin{eqnarray}
u_{V}\in U_{V}(1)\ , & \quad &g\in SU_{c}(3)\ , \nonumber  \\h(\xi
,g,u)\in SU_{c}(2)\ , & \quad & h_{\widetilde{V}}(\xi ,g,u) \in
\widetilde{U}_{V}(1)\ .
\end{eqnarray}
It is convenient to define the following differential form:
\begin{eqnarray}
\omega _{\mu }=i{V}^{\dagger }D_{\mu }{V}\quad {\rm with}\quad
D_{\mu }{V}=(\partial _{\mu }-ig_{s}G_{\mu }){V}\ ,
\end{eqnarray}
with $G_{\mu }=G_{\mu }^{m}T^{m}$ the gluon fields while $g_{s}$
is the strong coupling constant. $\omega $ transforms according
to:
\begin{eqnarray}
\omega _{\mu }&\rightarrow & h(\xi ,g,u)\omega _{\mu }h^{\dagger
}(\xi ,g,u)+i\,h(\xi ,g,u)\partial {_{\mu }}h^{\dagger }(\xi ,g,u)
\nonumber \\ &+& i\,h_{\widetilde{V}}(\xi ,g,u)\partial _{\mu
}h_{\widetilde{V}}^{\dagger }(\xi ,g,u) \ . \nonumber
\end{eqnarray}
We decompose $\omega _{\mu }$ into
\begin{eqnarray}
\omega _{\mu }^{\parallel }=2S^{a}{\rm Tr}\left[ S^{a}\omega _{\mu
}\right] \quad {\rm and}\quad \omega _{\mu }^{\perp }=2X^{i}{\rm
Tr}\left[ X^{i}\omega _{\mu }\right] \ ,
\end{eqnarray}
$S^{a}$ are the unbroken generators of $H$, while
$S^{1,2,3}=T^{1,2,3}$ and $S^{4}=\widetilde{B}\,/\sqrt{2}$.

The most generic two derivative kinetic Lagrangian for the
goldstone bosons is:
\begin{eqnarray}
L=f^{2}a_{1}{\rm Tr}\left[ \,\omega _{\mu }^{\perp }\omega ^{\mu
\perp }\, \right] +f^{2}a_{2}{\rm Tr}\left[ \,\omega _{\mu
}^{\perp }\,\right] {\rm Tr} \left[ \,\omega ^{\mu \perp
}\,\right] \ . \label{dt}
\end{eqnarray}
The double trace term is due to the absence of the condition for
the vanishing of the trace for the broken generator $X^{5}$. It
emerges naturally in the non linear realization framework at the
same order in derivative expansion with respect to the single
trace term. In the unitary gauge these two terms correspond to the
five gluon masses \cite{CDS}.
\subsection{The fermionic sector}
 \label{fermions} For the fermions it is convenient to define the
dressed fermion fields
\begin{eqnarray}
\widetilde{\psi}={V}^{\dagger }\psi \ ,  \label{mq}
\end{eqnarray}
transforming as $\widetilde{\psi}\rightarrow
h_{\widetilde{V}}(\xi,g,u)h(\xi ,g,u)\,\widetilde{\psi}$. $\psi$
has the ordinary quark transformations (i.e. is a Dirac spinor).
\begin{figure}[ht]
\vskip -1.5 cm
\begin{center}

\includegraphics[angle=0, height=3cm, width =3.in]{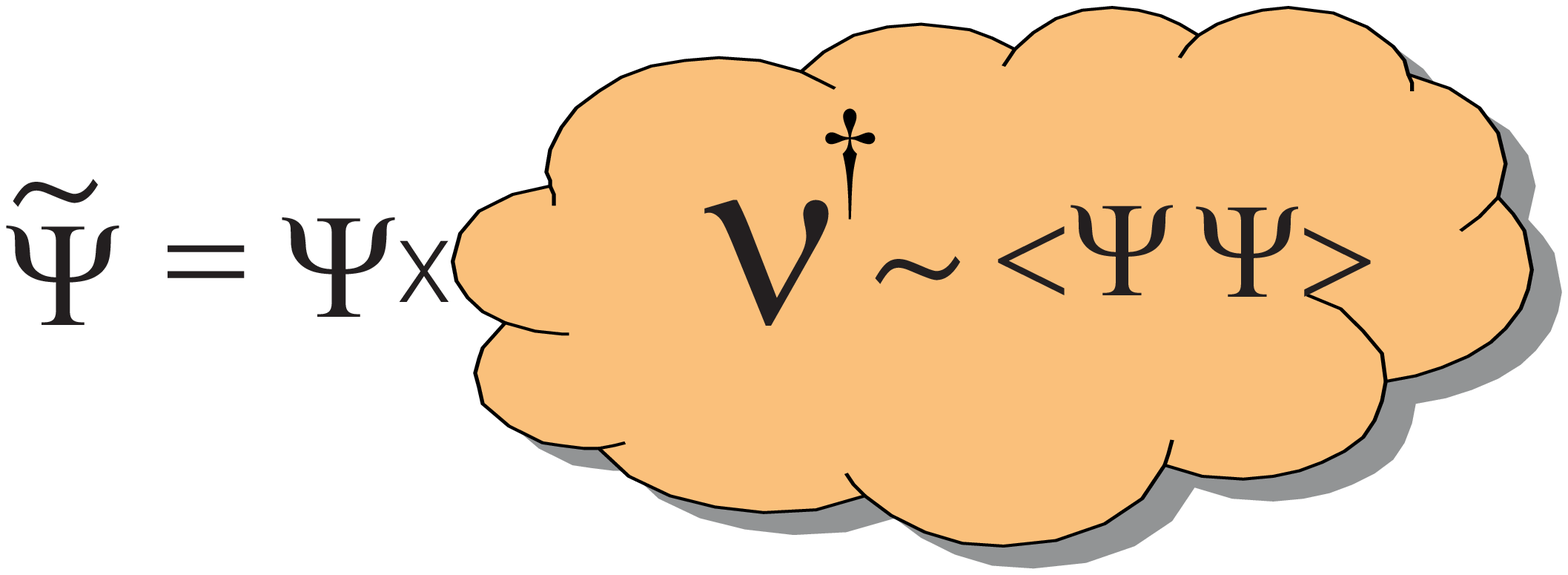}
\end{center}
\vskip -.6cm
\end {figure}
 Pictorially $\widetilde{\psi}$ can be viewed as a constituent
type field or alternatively as the bare quark field $\psi$
immersed in the diquark cloud represented by ${V}$. The non
linearly realized effective Lagrangian describing in medium
fermions, gluons and their self interactions, up to two
derivatives is:
\begin{eqnarray}
{L}&=&f^{2}a_{1}{\rm Tr}\left[ \,\omega _{\mu }^{\perp }\omega
^{\mu \perp }\,\right] +f^{2}a_{2}{\rm Tr}\left[ \,\omega _{\mu
}^{\perp }\,\right] {\rm Tr}\left[ \,\omega ^{\mu \perp }\,\right]
\nonumber \\ &+&b_{1}\overline{\widetilde{\psi }}i\gamma ^{\mu
}(\partial _{\mu }-i\omega _{\mu }^{\parallel })\widetilde{\psi
}+b_{2}\overline{\widetilde{\psi }} \gamma ^{\mu
}\omega _{\mu }^{\perp }\widetilde{\psi }  \nonumber \\
&+&m_{M}\overline{\widetilde{\psi }^{C}}_{i}\gamma
^{5}(iT^{2})\widetilde{\psi }_{j}\varepsilon ^{ij}+{\rm h.c.}\ ,
\end{eqnarray}
where $\widetilde{\psi }^{C}=i\gamma ^{2}\widetilde{\psi }^{\ast
}$, $i,j=1,2 $ are flavor indices and
\begin{eqnarray}
T^{2}=S^{2}=\frac{1}{2}\left(
\begin{array}{ll}
\sigma ^{2} & 0 \\
0 & 0
\end{array}
\right) \ .
\end{eqnarray}
Here $a_{1},~a_{2},~b_{1}$ and $b_{2}$ are real coefficients while
$m_{M}$ is complex.
{}From the last two terms, representing a Majorana mass term for
the quarks, we see that the massless degrees of freedom are the
$\psi _{a=3,i}$. The latter possesses the correct quantum numbers
to match the 't~Hooft anomaly conditions \cite{S}.

\subsection{The $SU_c(2)$ Glueball Lagrangian} The $SU_c(2)$ gauge
symmetry does not break spontaneously and confines. Calling $H$ a
mass dimension four composite field describing the scalar glueball
we can construct the following lagrangian \cite{OS2}:
\begin{eqnarray}
S_{G-ball}&=&\int
d^4x\left\{\frac{c}{2}\sqrt{b}\,H^{-\frac{3}{2}}\left[\partial^{0}
H
\partial^{0}H - v^2
\partial^iH
\partial^iH\right]  \right. \nonumber \\ &-& \left. \frac{b}{2}
H\log\left[\frac{H}{\hat{\Lambda}^4}\right] \right\} \ .
\label{G-ball}
\end{eqnarray}
This Lagrangian correctly encodes the underlying $SU_c(2)$ trace
anomaly. The glueballs move with the same velocity $v$ as the
underlying gluons in the 2SC color superconductor. $\hat{\Lambda}$
is related to the intrinsic scale associated with the $SU_c(2)$
theory and can be less than or of the order of few MeVs
\cite{Rischke:2000cn} \footnote{According to the present
normalization of the glueball field $\hat{\Lambda}^4$ is
$v\,\Lambda^4$ with $\Lambda$ the intrinsic scale of $SU_c(2)$
after the coordinates have been appropriately rescaled
\cite{Rischke:2000cn,{OS2}} to eliminate the $v$ dependence from
the action.} Once created, the light $SU_c(2)$ glueballs are
stable against strong interactions but not with respect to
electromagnetic processes \cite{OS2}. Indeed, the glueballs couple
to two photons via virtual quark loops.
\begin{eqnarray}
 \Gamma\left[h\rightarrow
\gamma\gamma\right] \approx 1.2\times 10^{-2}
\left[\frac{M_h}{1~{\rm MeV}}\right]^5~{\rm eV} \ ,
\end{eqnarray}
where $\alpha=e^2/4\pi \simeq 1/137$. {}For illustration purposes
we consider a glueball mass of the order of $1$~MeV which leads to
a decay time $\tau\sim~5.5\times~10^{-14}s$. This completes the
effective Lagrangian for the 2SC state which corresponds to the
Wigner-Weyl phase.

Using this Lagrangian one can estimate the $SU_c(2)$ glueball
melting temperature to be \cite{Sannino:2002re}:
\begin{eqnarray}
T_c \leq \sqrt[4]{\frac{90 {v}^3}{2\,e\, \pi^2}}\, {\hat{\Lambda}}
< T_{CSC} \ .
\end{eqnarray}
Where $T_{CSC}$ is the color superconductive transition
temperature.
\begin{figure}[ht]
\begin{center}
\includegraphics[angle=0,height = 2.5cm, width = 2in]{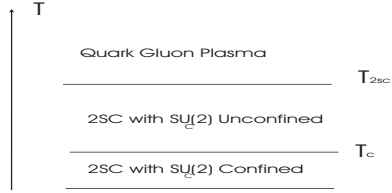}
\end {center}\caption[]{A zoom of the 2SC phases as function of temperature for fixed quark chemical potential.}
\end{figure}
The deconfining/confining $SU_c(2)$ phase transition within the
color superconductive phase is second order.

\section{Non Perturbative Exact Results: Anomaly Matching Conditions}
The superconductive phase for $N_{f}=2$ possesses the same global
symmetry group as the confined Wigner-Weyl phase. The ungapped
fermions have the correct global charges to match the t' Hooft
anomaly conditions as shown in \cite{S}. Specifically the
${SU(2)_{L/R}}^2\times U(1)_V$ global anomaly is correctly
reproduced in this phase due to the presence of the ungapped
fermions. This is so since a quark in the 2SC case is surrounded
by a diquark medium (i.e. $q \left< q q\right>$) and behaves as a
baryon.
\begin{eqnarray} \nonumber
\left(%
\begin{array}{c}
  u \\
  d \\
\end{array}%
\right)_{color = 3} \quad \sim \quad \left(%
\begin{array}{c}
  p \\
  n \\
\end{array}%
\right)\ .
\end{eqnarray}
The validity of the t'Hooft anomaly conditions at high matter
density have been investigated in \cite{S,{HSaS}}. A delicate part
of the proof presented in \cite{HSaS} is linked necessarily to the
infrared behavior of the anomalous three point function. In
particular one has to show the emergence of a singularity (i.e. a
pole structure). This pole is then interpreted as due to a
goldstone boson when chiral symmetry is spontaneously broken.

One might be worried that, since the chemical potential explicitly
breaks Lorentz invariance, the gapless (goldstone) pole may
disappear modifying the infrared structure of the three point
function. This is not possible. Thanks to the Nielsen and Chadha
theorem \cite{Nielsen}, not used in \cite{HSaS}, we know that
gapless excitations are always present when some symmetries break
spontaneously even in the absence of Lorentz
invariance\footnote{Under specific assumptions which are met when
Lorentz invariance is broken via the chemical potential.}. Since
the quark chemical potential is associated with the barionic
generator which commutes with all of the non abelian global
generators the number of goldstone bosons must be larger or equal
to the number of broken generators. Besides all of the goldstones
must have linear dispersion relations (i.e. are type I
\cite{Nielsen}). This fact not only guarantees the presence of
gapless excitations (justifying the analysis made in \cite{HSaS}
on the infrared behavior of the form factors) but demonstrates
that the pole structure due to the gapless excitations needed to
saturate the triangle anomaly is identical to the zero quark
chemical potential one in the infrared.

It is also interesting to note that the explicit dependence on the
quark chemical potential is communicated to the goldstone
excitations via the coefficients of the effective Lagrangian (see
\cite{Schafer:2003vz} for a review). {}For example $F_{\pi}$ is
proportional to $\mu$ in the high chemical potential limit and the
low energy effective theory is a good expansion in the number of
derivatives which allows to consistently incorporate in the theory
the Wess-Zumino-Witten term \cite{S} and its corrections.

The validity of the anomaly matching conditions have far reaching
consequences. Indeed, in the three flavor case, the conditions
require the goldstone phase to be present in the hadronic as well
as in the color superconductive phase supporting the quark-hadron
continuity scenario \cite{Schafer:1998ef}. At very high quark
chemical potential the effective field theory of low energy modes
(not to be confused with the goldstone excitations) has positive
Euclidean path integral measure \cite{Hong:2003zq}. In this limit
the CFL is also shown to be the preferred phase with the aid of
the anomaly conditions. Since the fermionic theory has positive
measure only at asymptotically high densities one cannot use this
fact to show that the CFL is the preferred phase for moderate
chemical potentials. This is possible using the anomaly
constraints.

While the anomaly matching conditions are still in force at
nonzero quark chemical potential \cite{S} the {\it persistent
mass} condition \cite{Preskill:1981sr} ceases to be valid. Indeed
a phase transition, as function of the strange quark mass, between
the CFL and the 2SC phases occurs.

We recall that we can saturate the t'Hooft anomaly conditions
either with massless fermionic degrees of freedom or with gapless
bosonic excitations. However in absence of Lorentz covariance the
bosonic excitations are not restricted to be fluctuations related
to scalar condensates but may be associated, for example, to
vector condensates \cite{Sannino:2001fd}.



\begin{acknowledgments}
I thank A.D. Jackson for stimulating collaboration on some of the
recent topics presented here and for careful reading of the
manuscript. I also thank R. Casalbuoni, P.H. Damgaard, Z. Duan, M.
Harada, D.K. Hong, S.D. Hsu, R. Marotta, A.~M\'{o}csy, F.
Pezzella, J. Schechter, M. Shifman, K. Splittorff and K. Tuominen
for their valuable collaboration and/or discussions on some of the
topics I presented here. K. Tuominen is thanked also for careful
reading of the manuscript.
\end{acknowledgments}



%




\end{document}